\documentclass[prl,twocolumn,superscriptaddress]{revtex4-1}
\usepackage{}
\usepackage{amssymb}
\usepackage{amsfonts}
\usepackage{bbm}
\usepackage{mathrsfs}
\usepackage{graphics,graphicx,epsfig,bm,amsmath,amsthm,amssymb}
\usepackage{bm}
\usepackage{bbm}
\usepackage{longtable}
\usepackage{multirow}
\usepackage{array}
\usepackage{color}
\usepackage[usenames,dvipsnames]{xcolor}
\usepackage{amsmath}
\usepackage{verbatim}
\usepackage{float}
\usepackage[a4paper,colorlinks=true,
linkcolor=blue,citecolor=blue,
pdfauthor={ },
pdftitle={ },
pdfsubject={ },
pdfkeywords={ }]{hyperref}

\begin{document}

\title{Quantum state tomography of a single electron spin in diamond with Wigner reconstruction}

\author{Bing Chen}
\affiliation{School of Electronic Science and Applied Physics,Hefei University of Technology, Hefei,
Anhui 230009, China}
\affiliation{State Key Laboratory of Quantum Optics and Quantum Optics Devices, 
\\Shanxi University, Taiyuan, 030006,China}%

\author{Jianpei Geng}
\affiliation{School of Electronic Science and Applied Physics,Hefei University of Technology, Hefei,
Anhui 230009, China}

\author{Feifei Zhou}
\affiliation{School of Electronic Science and Applied Physics,Hefei University of Technology, Hefei,
Anhui 230009, China}

\author{Lingling Song}
\affiliation{School of Electronic Science and Applied Physics,Hefei University of Technology, Hefei,
Anhui 230009, China}

\author{Heng Shen}
\email{heng.shen@physics.ox.ac.uk}
\affiliation{Clarendon Laboratory, University of Oxford, Parks Road, Oxford, OX1 3PU, UK}

\author{Nanyang Xu}
\email{nyxu@hfut.edu.cn}
\affiliation{School of Electronic Science and Applied Physics,Hefei University of Technology, Hefei,
Anhui 230009, China}%
%\end{affiliations}

%\date{\today}

\begin{abstract}
We present the experimental reconstruction of the Wigner function of an individual electronic spin qubit associated with a nitrogen-vacancy ($\text{NV}$) center in diamond at room temperature. This spherical Wigner function contains the same information as the density matrix for arbitrary spin systems. As an example, we exactly characterize the quantum state of a single qubit undergoing a nearly pure dephasing process by Wigner function. The fidelities and purities during this process are extracted from the experimental reconstructed Wigner functions, whose dynamics agree with the theoretical prediction. Our method can be applied to multi-qubit systems for measuring the Wigner function of their collective spin state. 
\end{abstract}
\date{\today}%
%\pacs{Valid PACS appear here}

\maketitle
Quantum state tomography of a system from measurements is an important topic in the emerging field of quantum technology.
Through full state reconstruction, one can estimate the properties of quantum systems such as entanglement and purity,
and furthermore determine their potential application in fields of quantum metrology\cite{C. L. Degen,T. Rendler,T. Unden,Appel,Vladan,Gross}, quantum simulation\cite{Friedenauer, Ben,Kim}
and quantum computation\cite{P. Neumann,T. H. Taminiau,J. Wagenaar,F. Kong,K. Xu,Leibfried,Benhelm}. The most common method for full characterization of any quantum state is the density matrix reconstruction. However, 
%with the number of qubits increasing, 
as the number of qubits increases, the density matrix reconstruction with maximum likelihood estimation\cite{Hradil} in practice becomes problematic. 
%The spherical Wigner function defines the quasiprobability distribution for systems of arbitrary angular momentum in spherical phase space.
% while for continuous variable systems, Wigner functions are widely utilized to fully characterize the quantum state. %As a counterpart of density matrix reconstruction, Wigner function is convenient. 

As a counterpart of density matrix reconstruction, the Wigner function was originally proposed for describing quantum systems with continuous degrees of freedom, for instance the harmonic-oscillator
phase-space description of electromagnetic fields\cite{Lvovsky}. Many efforts have been made to generalize the method of the Wigner function to quantum systems with a finite-dimensional Hilbert space\cite{Agarwal,Schmied,Wootters,Leonhardt,Vourdas,Miquel,Robert McConnell,Treutlein,Tilma,IBM,JLi}, such as systems of arbitrary angular momentum in spherical phase space. However, a complete reconstruction of the Wigner function in solid-state spin systems has not been experimentally realized.

%In this letter, we apply the method proposed in reference 18 to characterize the states of an electron spin of a nitrogen-vacancy (NV) center in diamond with Wigner functions. We experimentally construct the complete and continuous Wigner functions for the states of the electron spin undergoing a dephasing process. The fidelities with an ideal initial state and purities of the states are extracted from the experimental Wigner functions. The decrease of the fidelities and purities are in accordance with the dephasing time measured by a Ramsey sequence. 
Generally, as for spin system consisting of $\text{N}$ atoms, with each atom representing
a pseudo-spin-1/2 subsystem, the corresponding Wigner function is given as\cite{Schmied,Robert McConnell,Agarwal}
\begin{equation}\label{W_general}
\text{W}(\theta,\phi)=\sqrt{\frac{2}{\pi}}\sum_{k=0}^{2j}\sum_{q=-k}^{k}\text{Y}_{kq}(\theta,\phi)\rho_{kq},
\end{equation}
where $j = N/2$ is the total spin length, and $\text{Y}_{kq}$ are the usual spherical harmonics. $\theta$ is the polar angle measured
from the +z-axis, and $\phi$ is the azimuthal angle around the z-axis. Here, the density matrix $\rho$ is transformed from j-space
(the Dicke representation $\rho_{mm'}=\left \langle m \left | \rho \right |m' \right \rangle$ of the density matrix) to k-space (the
spherical harmonic decomposition $\rho_{kq}$ by $\rho_{kq}=\sum_{m=-j}^{j}\sum_{m'=-j}^{j}\rho_{mm'}t^{jmm'}_{kq}$ with multipole
operator-related coefficient $t^{jmm'}_{kq}=(-1)^{j-m-q}\left \langle j,m;j,-m' | k,q \right \rangle$,
where $\left \langle j,m;j,-m' | k,q \right \rangle$ is Clebsch Gordan coefficient. This Wigner function contains the same information as the
density matrix for any spin- j system, further, the expectation value of the angular momentum vector is proportional to the center of mass of the Wigner function, $\left \langle \text{J}_i \right \rangle=\sqrt{\frac{j(j+1)(2j+1)}{4\pi}}\int_{0}^{\pi}\sin(\theta)d\theta\int_{0}^{2\pi}d\phi f_i(\theta,\phi)\text{W}(\theta,\phi)$, where $i=(x,y,z)$ while $f_i(\theta,\phi)=\left \{
\sin\theta\cos\phi,\sin\theta\sin\phi,\cos\theta \right \}$.

In this letter, we experimentally reconstruct the complete and continuous Wigner functions for the states of the electron spin of an $\text{NV}$ center in diamond. Experiencing a nearly pure dephasing process, it is demonstrated that 
the decay of the fidelities and purities extracted from the reconstructed Wigner functions is in accordance with the dephasing time measured by a Ramsey sequence\cite{R. Hanson,L. Childress}. Additionally, the minimum value of the Wigner function ($\text{W}_{min}$) is presented, which increases from negativity to positivity with the electron spin dephasing into a more and more mixed state. The negativity completely vanishes when the purity of the spin state extracted from the Wigner function is less than $2/3$ which is the main conclusion of Ref\cite{JLi}. However, due to the limited slices of Wigner function, the authors obtained the purity from typical density matrix reconstruction in that work.

\begin{figure}[htb]
\centering
\includegraphics[width=0.45\textwidth]{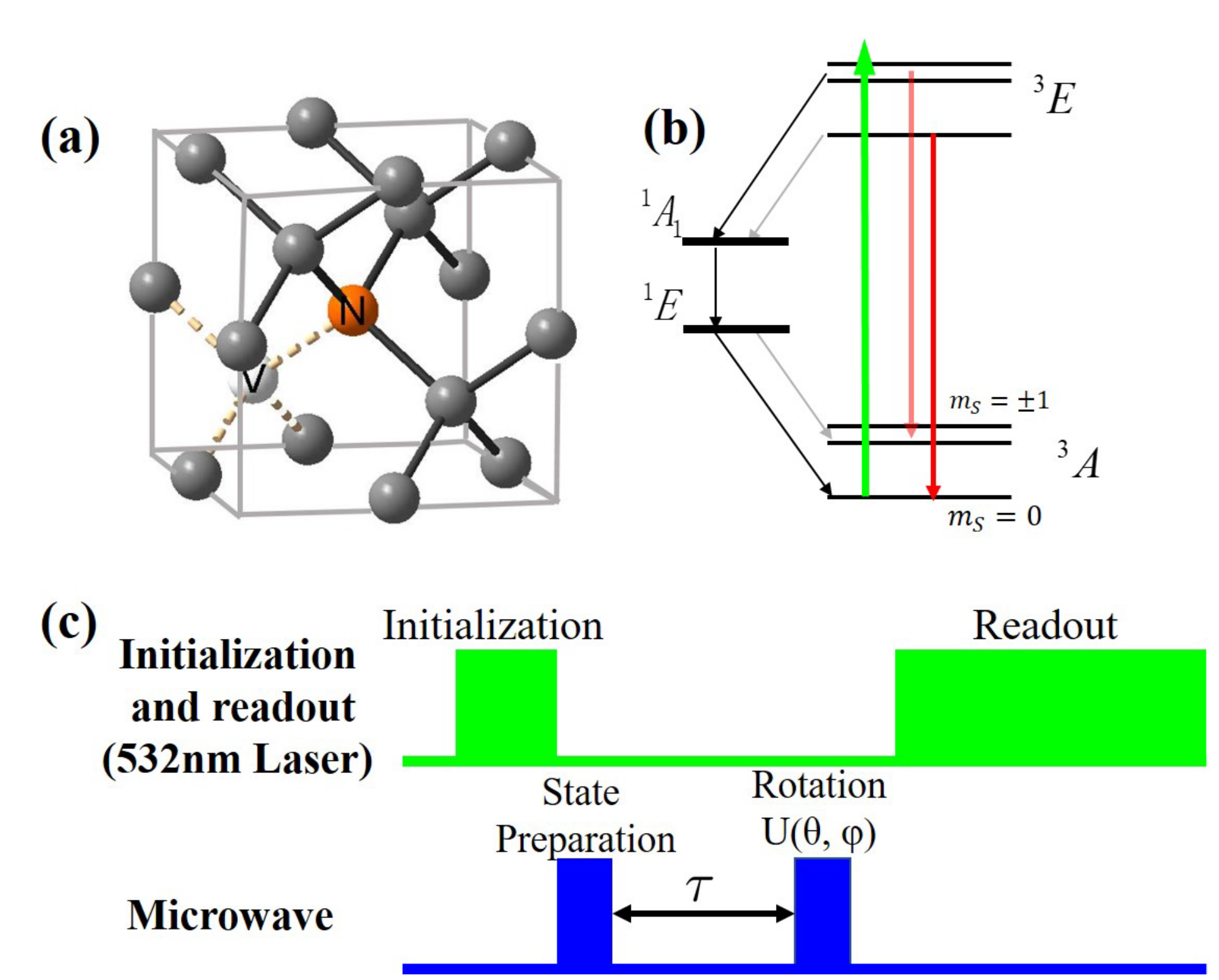}
\caption{\label{Fig 1}(color online). (a) Atomic structure of NV center in diamond. The $\text{NV}$ center in diamond which consists of a substitutional nitrogen atom (N)associated with a vacancy (V) in an adjacent lattice site of the diamond matrix\cite{Doherty}, has $\text{C}_{\text{3v}}$ symmetry. (b) Scheme of energy levels of the NV center electron spin. Its ground state ($^{3}\text{A}_2$) and excited state
($^{3}\text{E}$) are both spin triplet, and the transition between the two states is corresponding to the zero-phonon line(ZPL) at $637\text{nm}$ ($1.945\text{eV}$). The ground state ($^{3}\text{A}_2$) is a spin triplet with zero-field splitting of $2.87\text{GHz}$ between $m_s=0$ and $m_s=\pm 1$ states. The excited states($^{3}\text{E}$) is governed by spin-orbit and spin-spin interactions, split by $1.43\text{GHz}$ between $m_s=0$ and $m_s=\pm 1$ states, all excited states spin levels (spin quantum number $m_s=0, \pm 1$) exhibit spontaneous decay by photon emission. (c) The laser and microwave pulse sequence for the measurement of the Wigner function. The second microwave pulse corresponds to the unitary operation $U(\theta,\phi)$=$e^{-i\theta/2(\cos\phi\sigma_x+\sin\phi\sigma_y)}$. $\tau$ is the dephasing time.}
\end{figure}

The quantum-mechanical state of any two-level system can be expressed as a $2\times 2$ density matrix $\rho=\frac{1}{2}(\mathbb{I}+\mathbf{r}\cdot\hat{\sigma})$ with the vectors
$\mathbf{r} = (x, y, z)\in \mathbb{R}^3$ ($\left \| \mathbf{r} \right \|=\sqrt{x^2+y^2+z^2}\leqslant 1$) and
Pauli operators $\hat{\mathbf{\sigma}} = (\hat{\sigma}_x, \hat{\sigma}_y, \hat{\sigma}_z)$. To visualize the spin state pointing on the
surface of Bloch Sphere, $\mathbf{r}$ is typically represented as $\mathbf{r} = r\cdot(\sin\epsilon\cos\eta,\sin\epsilon\sin\eta,\cos\eta)$
where $\epsilon$ and $\eta$ are the polar and azimuthal angle, respectively. Inserting this density matrix of single qubit to Eq.\ref{W_general},
one can achieve the theoretical Wigner function for single qubit as follows,
\begin{equation}\label{W_single}
\text{W}(\theta,\phi)=\frac{1-\sqrt{3}r[\sin\epsilon\sin\theta\sin(\eta-\phi)+\cos\epsilon\cos\theta]}{2\pi^2}.
\end{equation}

\begin{figure}
\centering
\includegraphics[width=0.45\textwidth]{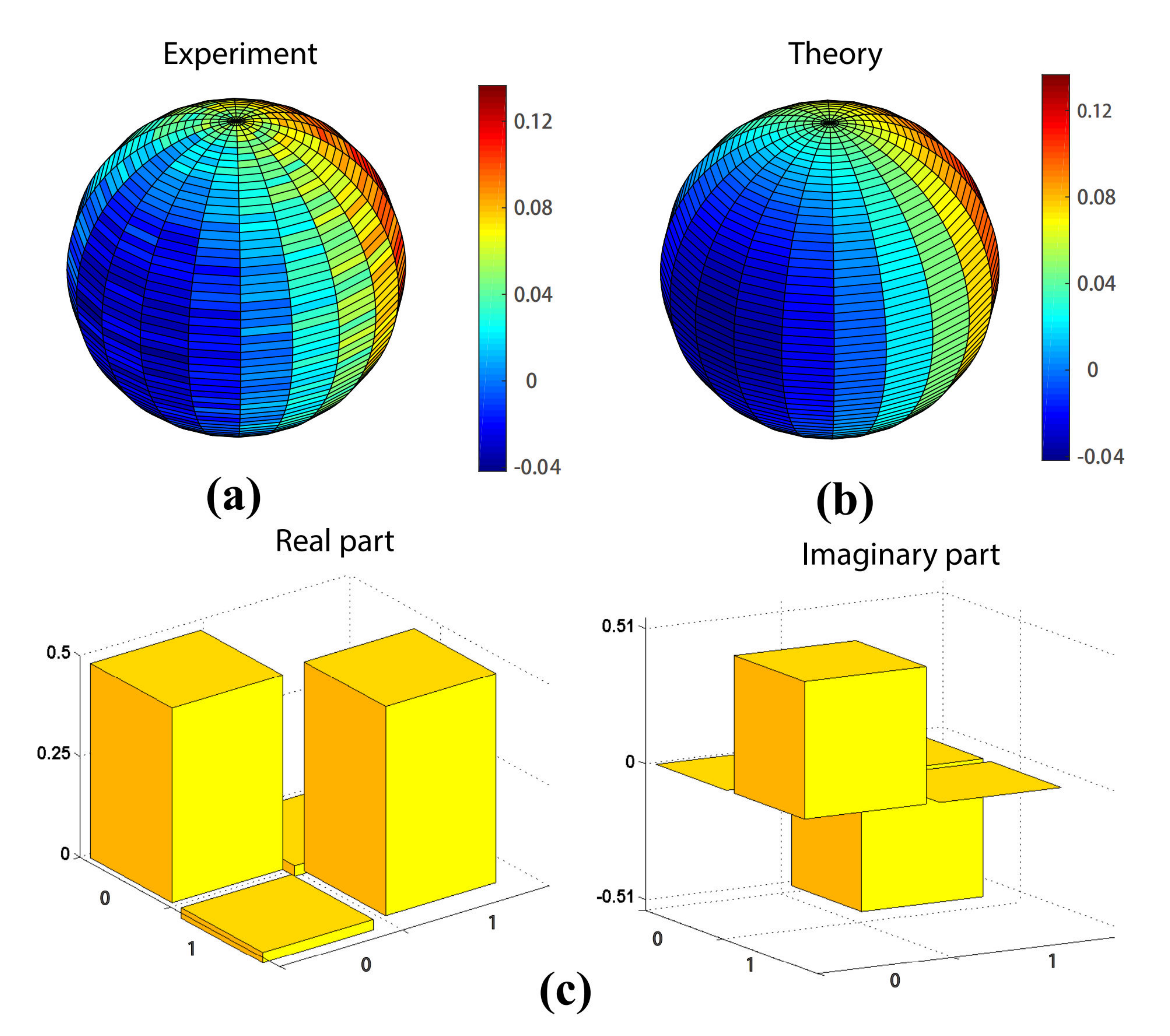}
\caption{\label{Fig 2}(color online). Experimental (a) and theoretical (b) spherical Wigner function W($\theta,\phi$) of a qubit in state of 
$\left|\text{y}\right \rangle =(\left |0\right \rangle+i\left |1\right \rangle)/{\sqrt{2}}$ reconstructed on the curved Bloch sphere. 
$\theta$ is the polar angle measured from the +z-axis, and $\phi$ is the azimuthal angle around the z-axis. (c) Real and imaginary parts, respectively, of the reconstructed density matrix elements of the qubit state. Each data point has been averaged $10^6$ repetitions. The error bars account for the statistical error associated with the photon counting.}
\end{figure}

In general, due to connecting to the environment bath spin states experience a complicated decoherence process that contains both dissipative
and dephasing dynamics. To clarify the dynamics of Wigner function of the single qubit, we experimentally prepare the spin state along y-axis as
$\left |  \text{y}\right \rangle =\frac{\left |0\right \rangle+i\left |1\right \rangle}{\sqrt{2}}$, i.e. $\epsilon=\pi/2, \eta=\pi/2$,
which mainly suffers from the dephasing dynamics.

In the experiment, we use a purpose-built confocal microscopy to address and detect single nitrogen-vacancy centers in a type-IIa, single-crystal synthetic diamond sample (Element Six)\cite{Doherty}. The atomic structure and energy levels of the NV center in diamond are schematically shown in Fig.\ref{Fig 1}a and Fig.\ref{Fig 1}b, respectively. By applying a laser pulse of $532$~nm wavelength with the assistance of intersystem crossing (ISC) transitions, the spin state can be polarized into $m_s=0$ in the ground state($^{3}\text{A}_2$). This process can be utilized to initialize and read out the spin state of the NV center. The fluorescence photons are detected by using the single photon counting module (SPCM). Additionally, a small permanent magnet in the vicinity of the diamond (magnetic field $B\approx 520\text{G}$) that is aligned parallel to the symmetry axis of the nitrogen vacancy center splits the $m_s=\pm1$ spin levels. With this magnetic field, the $^{14}N$ nuclear spin of the NV center can be also polarized with the laser pulse\cite{nuclear pol}. It is observed in the optically detected magnetic resonance (ODMR) spectra that the nuclear spin polarization is higher than $98\%$. We encode $m_s=-1$ and $m_s=0$  in $^{3}\text{A}_2$ as $|0\rangle$ and $|1\rangle$ of the electron spin qubit. The state of the qubit can be manipulated with
microwave pulses(1404.3MHz), while the spin level $m_s=1$ remains idle due to large detuning.

To make the Wigner function Eq. \ref{W_single} more closely linked to the experimental implementation, we rewrite it as
\begin{equation}\label{W_measure}
 \begin{aligned}
  \text{W}(\theta,\phi)&=\sqrt{\frac{2}{\pi}}\sum_{m=-j}^{j}p_m(\theta,\phi)R_{mj},\\
   R_{mj}&=\sum_{k=0}^{2j}\sqrt{\frac{2k+1}{4\pi}}t^{jmm}_{k0}\\
   &=\frac{(-1)^{j-m}}{\sqrt{4\pi}}\sum_{k=0}^{2j}(2k+1)\bigl(\begin{smallmatrix}
 j& j & k\\
 m& -m & 0
\end{smallmatrix}\bigr),
 \end{aligned}
\end{equation}
where $p_m(\theta,\phi)$ is the spin projection probabilities projected along a specific quantization axis $(\theta,\phi)$ with $\sum_{m=-j}^{j}p_m=1$.
And the coefficient $R_{mj}$ is written in terms of a Wigner $3j$-symbol. Based on the experimental equation, we can reconstruct the Wigner function of the spin state from experimental result.

For each experimental cycle of Wigner function reconstruction as shown in Fig.\ref{Fig 1}c, we start the sequence with $1\mu s$ of laser illumination to polarize the nitrogen-vacancy electron spin and nearby nuclear spins into state $|1\rangle$. Then the qubit is prepared into state $(|0\rangle+i|1\rangle)/\sqrt{2}$ by a $\pi/2$ microwave pulse. An idle time of $\tau$ is followed, during which the qubit dephases into a mixed state. For the measurement of the Wigner function of the state, we apply a second microwave pulse with phase $\phi$, Rabi frequency $\Omega_W$, and duration $\theta/\Omega_W$. The second microwave pulse thus corresponds to the unitary operation $e^{-i\theta/2(\cos\phi\sigma_x+\sin\phi\sigma_y)}$. After the microwave pulse, another $1\mu$s laser pulse is applied and fluorescence emission is detected by the single photon counting module and be normalized into the population of state $|1\rangle$ and $|0\rangle$. We can obtain the expectation of $p_m(\theta,\phi)$ and reconstruct exactly the experimental Wigner function basing on Eq.\ref{W_measure}.
To reconstruct the Wigner function, we vary $\theta$ from 0 to $\pi$ with a step of $\pi/60$ and $\phi$ from 0 to $2\pi$ with a step of $\pi/10$. The variation of $\theta$ and $\phi$ is realized by varying the duration and phase of the microwave pulse, which is generated from an IQ-modulation system where we use an arbitrary waveform generator (Tektronix AWG510) to synthesize different frequencies and phases. The generated microwave pulse is passed through a switch, amplified by a power amplifier, and delivered by an impedance-matched coplanar waveguide (CPW) before being applied on the qubit. 

\begin{figure}
\centering
\includegraphics[width=0.5\textwidth]{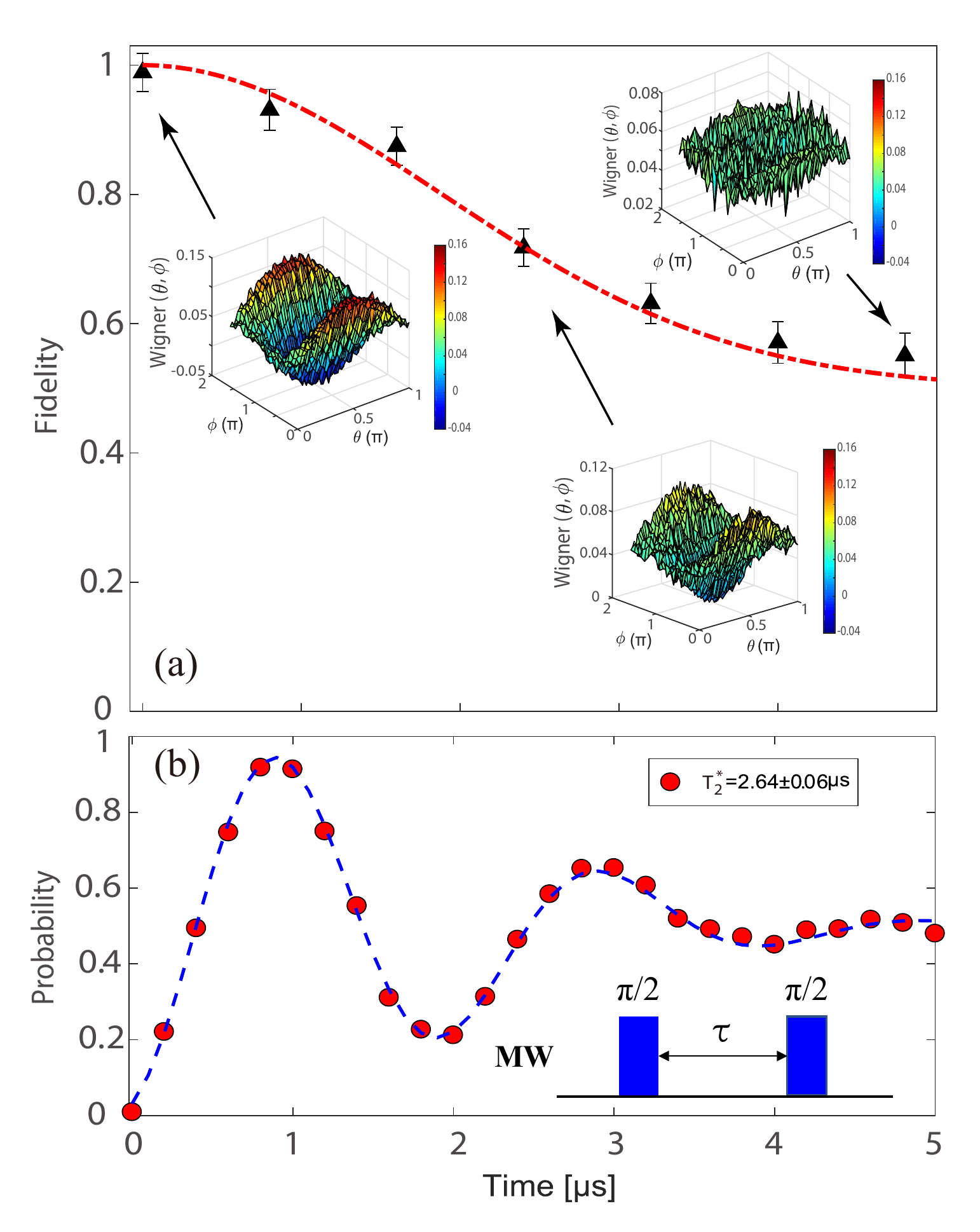}
\caption{\label{Fig 3}(color online). (a) Dynamics of the Wigner function-extracted fidelity in a nearly pure dephasing process. The black triangle represents the fidelity extracted from experimental Wigner function. The dashed line is the theoretical prediction. The three insets are the surface-plotted Wigner function ($W_\tau$) at time $\tau=0\mu$s, $2.4\mu$s and 4.8$\mu$s. (b) The Ramsey oscillation of the electron spin coherence. The data were taken with the microwave detuning of 0.5 MHz by varying the temporal separation between the two microwave $\pi/2$ pulses. The Ramsey signal(red circle) was fitted to exp$[-(\tau/T_2^*)^2]\cos(2\pi ft)$ (blue line)\cite{R. Hanson, L. Childress} where $f$ values correspond to the microwave detuning, obtained $T_2^*=2.64\pm 0.06\mu$s. The inset is the experimental microwave pulse sequence of the electron-spin free precession. Each data point has been averaged $10^6$ repetitions. The error bars account for the statistical error associated with the photon counting.}
\end{figure}

Fig.\ref{Fig 2}a presents the spherical Wigner function for the state $\rho_0=|\text{y}\rangle\langle \text{y}|$ experimentally prepared at the initial time. The color on the Bloch sphere indicates the value of W($\theta,\phi$). As comparison, Fig.\ref{Fig 2}b shows the theoretical Wigner function based on Eq.\ref{W_single} which agrees with the experimental Wigner function in Fig.\ref{Fig 2}a. The density matrix of the spin state via standard quantum state tomography is shown in Fig.\ref{Fig 2}c, with $\rho_{0,\text{exp}}=\bigl(\begin{smallmatrix}
 0.482\pm 0.014&  -0.026\pm 0.018 - i(0.518\pm 0.013) \\
  -0.026\pm 0.018 + i(0.516\pm 0.013)& 0.518\pm 0.014
\end{smallmatrix}\bigr)$.

\begin{figure}
\centering
\includegraphics[width=0.45\textwidth]{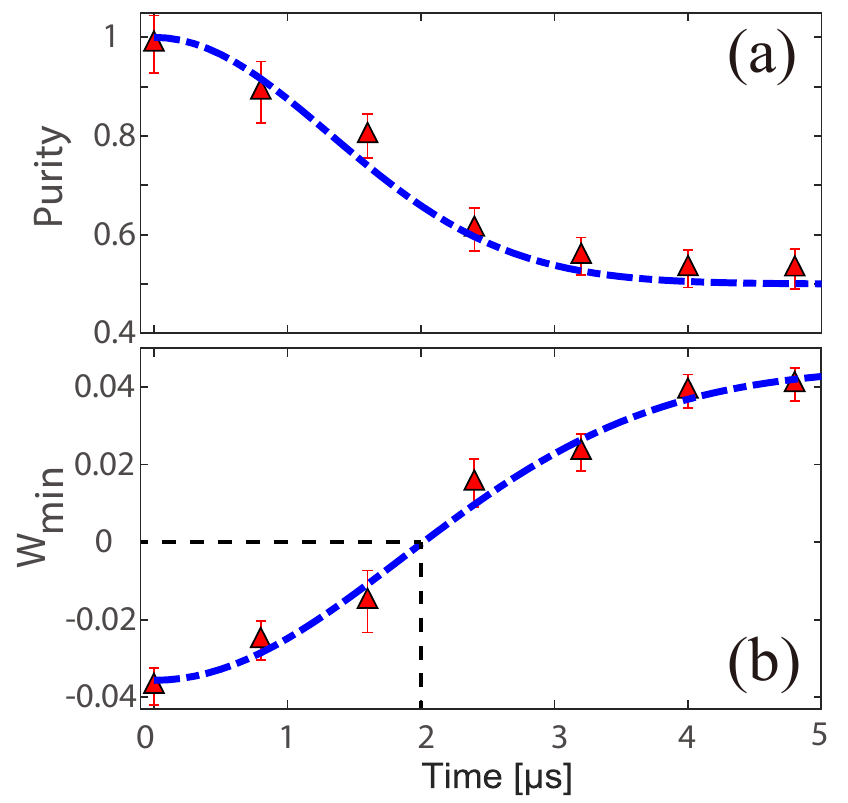}
\caption{\label{Fig 4}(color online). (a) Dynamics of the Wigner function-extracted purity in a nearly pure dephasing process. The red triangle is the extracted purity, and the blue dashed line is theoretical prediction. (b) Measured $\text{W}_{min}$ in a nearly pure dephasing process. The red triangle is the experimental data, and the blue dashed line represents a theoretical curve  $\frac{1-\sqrt{3}\exp[-2(\tau/T_2^*)^2]}{2\pi^2}$ derived from Eq.\ref{W_single}. Each data point has been averaged $10^6$ repetitions. The error bars account for the statistical error associated with the photon counting.}
\end{figure}

As the state evolves in the dephasing process, the state becomes more and more mixed. Starting from the ideal initial state $\rho_0$, the state should evolve as $\rho(\tau)=\frac{1}{2}\bigl(\begin{smallmatrix}
 1&  -i\text{exp}[-(\tau/T_2^*)^2] \\
  i\text{exp}[-(\tau/T_2^*)^2]& 1
\end{smallmatrix}\bigr)$\cite{Lecture}, where $T_2^*$ is the dephasing time. We apply Ramsey sequence on the electron spin to measure $T_2^*$. The Ramsey sequence is $\pi/2$-$\tau$-$\pi/2$ which is shown in the inset of Fig.\ref{Fig 3}b). The dephasing time $T_2^*=2.64\pm 0.06\mu$s can be obtained by fitting of the experimental Ramsey signal shown as the red circles in Fig.\ref{Fig 3}b. We use fidelity and purity of the state to evaluate this dephasing dynamics. The fidelity and purity of the state $\rho(\tau)$ are defined as $F=(\text{Tr}(\sqrt{\sqrt{\rho_{0}}\rho{(\tau)}\sqrt{\rho_{0}}}))^2$ and $P=\text{Tr}(\rho{(\tau)}^2)$, which gives $F=0.5+0.5\text{exp}[-(\tau/T_2^*)^2]$ and $P=0.5+0.5\text{exp}[-2(\tau/T_2^*)^2]$ as theoretical prediction. In the language of Wigner functions, they can be expressed as $F=2\pi\int_{0}^{\pi}sin (\theta)d\theta\int_{0}^{2\pi}d\phi \text{W}_\tau(\theta,\phi)\text{W}_0(\theta,\phi)$ and $P=4\pi\int_{0}^{\pi}sin (\theta)d\theta\int_{0}^{2\pi}d\phi \text{W}_\tau^2(\theta,\phi)$, where $\text{W}_\tau$ and $\text{W}_0$ are the Wigner functions corresponding to $\rho(\tau)$ and $\rho_0$, respectively. Fig.\ref{Fig 3}a and Fig.\ref{Fig 4}a show the fidelity and purity extracted from the experimental Wigner function $W_\tau$ which are in good agreement with the theoretical predictions.

Further, we find that the Wigner function of the prepared initial state has negative region and $\text{W}_{min}$ increases gradually. As expected as the simulation (Eq.\ref{W_single}) (Fig.\ref{Fig 4}b), the negativity of the Wigner function completely vanishes around $2.0\mu s$ when the purity of the spin state extracted from the Wigner function is less than $2/3$, in agreement with the main observation in Ref\cite{JLi}. 

In conclusion, we report the experimental reconstruction of the spherical Wigner function of a single qubit state for an $\text{NV}$ center in a bulk diamond. In a nearly pure dephasing process, Wigner functions at different time are measured to extract the dynamical information of the spin states. We present the dynamics of the Wigner function-extracted fidelity and purity whose behavior agrees with the theoretical prediction. Our method can be applied to multi-spin systems for quantum state tomography instead of density matrix reconstruction, which is problematic
in large spin systems.

The authors are grateful to Jie Li and Yanqiang Guo for fruitful discussions. The authors acknowledge financial support by the National Natural Science Foundation
of China (grant No. 11604069, 61376128), by the National Key R\&D Program of China(Grants No. 2018YFA0306600), the Anhui Natural Science Foundation (grant No. 1708085QA09), the Program of State Key Laboratory of Quantum Optics and Quantum Optics Devices(No.KF201802) and the Fundamental Research Funds for the Central Universities. H. Shen acknowledges the financial support from the Royal Society Newton International Fellowship 
(NF170876) of UK.

\end{document}